\documentclass[sigconf]{acmart} 

\usepackage{colortbl}
\usepackage{multirow}
\usepackage{graphicx}
\usepackage{subfigure}
\usepackage{caption}
\usepackage{enumitem}

\AtBeginDocument{%
  \providecommand\BibTeX{{%
    \normalfont B\kern-0.5em{\scshape i\kern-0.25em b}\kern-0.8em\TeX}}}

\copyrightyear{2024} 
\acmYear{2024} 
\setcopyright{rightsretained} 
\acmConference[WWW '24 Companion]{Companion Proceedings of the ACM Web Conference 2024}{May 13--17, 2024}{Singapore, Singapore}
\acmBooktitle{Companion Proceedings of the ACM Web Conference 2024 (WWW '24 Companion), May 13--17, 2024, Singapore, Singapore}
\acmDOI{10.1145/3589335.3648332}
\acmISBN{979-8-4007-0172-6/24/05}

\begin{document}

\title{Hierarchical Query Classification in E-commerce Search}


\author{Bing He}
\email{bhe46@gatech.edu}
\affiliation{%
  \institution{Georgia Institute of Technology}
  \city{Atlanta}
  \state{GA}
  \country{USA}
}

\author{Sreyashi Nag}
\email{sreyanag@amazon.com}
\affiliation{%
  \institution{Amazon}
  \city{Palo Alto}
  \state{CA}
  \country{USA}}

\author{Limeng Cui}
\email{culimeng@amazon.com}
\affiliation{%
  \institution{Amazon}
  \city{Palo Alto}
  \state{CA}
  \country{USA}}

\author{Suhang Wang}
\email{szw494@psu.edu}
\affiliation{%
 \institution{The Pennsylvania State University}
 \city{University Park}
 \state{PA}
 \country{USA}}

\author{Zheng Li}
\email{amzzhe@amazon.com}
\affiliation{%
  \institution{Amazon}
  \city{Palo Alto}
  \state{CA}
  \country{USA}}

\author{Rahul Goutam}
\email{rgoutam@amazon.com}
\affiliation{%
  \institution{Amazon}
  \city{Palo Alto}
  \state{CA}
  \country{USA}}

\author{Zhen Li}
\email{amzzhn@amazon.com}
\affiliation{%
  \institution{Amazon}
  \city{Palo Alto}
  \state{CA}
  \country{USA}}

\author{Haiyang Zhang}
\email{hhaiz@amazon.com}
\affiliation{%
  \institution{Amazon}
  \city{Palo Alto}
  \state{CA}
  \country{USA}}


\begin{abstract}

E-commerce platforms typically store and structure product information and search data in a hierarchy. 
Efficiently categorizing user search queries into a similar hierarchical structure is paramount in enhancing user experience on e-commerce platforms as well as news curation and academic research. 
The significance of this task is amplified when dealing with sensitive query categorization or critical information dissemination, where inaccuracies can lead to considerable negative impacts. 
The inherent complexity of hierarchical query classification is compounded by two primary challenges: (1) the pronounced class imbalance that skews towards dominant categories, and (2) the inherent brevity and ambiguity of search queries that hinder accurate classification.

To address these challenges, we introduce a novel framework that leverages hierarchical information through 
(i) enhanced representation learning that utilizes the contrastive loss to discern fine-grained instance relationships within the hierarchy, called ``instance hierarchy'', 
and (ii) a nuanced hierarchical classification loss that attends to the intrinsic label taxonomy, named ``label hierarchy''.
Additionally, based on our observation that certain unlabeled queries share typographical similarities with labeled queries, we propose a neighborhood-aware sampling technique to intelligently select these unlabeled queries to boost the classification performance.
Extensive experiments demonstrate that our proposed method is better than state-of-the-art (SOTA) on the proprietary Amazon dataset, and comparable to SOTA on the public datasets of Web of Science and RCV1-V2.
These results underscore the efficacy of our proposed solution, and pave the path toward the next generation of hierarchy-aware query classification systems.

\end{abstract}





\settopmatter{printfolios=true} 

\maketitle

\section{Introduction}

Hierarchical query classification is a vital task in the domain of e-commerce and search, playing a crucial role in driving customer obsession~\cite{chuang2002towards}. As users interact with online services, they input various queries to search for products, services, or information. 
Accurately classifying these queries is pivotal in ensuring that users are presented with the most relevant and valuable results. 
One significant application of the hierarchical query classifier in industry is categorizing sensitive queries that follow a predefined hierarchy in e-commerce. 
For example, given a query, it can be classified as harmful, adult-oriented, or non-sensitive products (Here, for illustration, we define these categories as parent categories). Furthermore, for harmful products, there are two child categories: self-harm and harm to others. The child categories for the adult-oriented category can be adult products and adult content.
Since these queries contain offensive content or pertain to unregulated goods, and different categories need to be handled differently, mis-classification of such queries can lead to unpleasant or even detrimental user experiences, potentially damaging a service's reputation and user trust. Moreover, presenting inappropriate or restricted content could lead to legal ramifications for the service. 
Hence, building an accurate hierarchical query classification framework is of paramount importance, not just for user satisfaction, but also for the overall compliance and integrity of the service~\cite{shen2009product}.

Various machine learning techniques are employed to identify the appropriate hierarchical category for each query~\cite{zhu2023hcl4qc, chuang2002towards, zhou2017survey}, and sense the context and nuances each query presents~\cite{ye2020zero}. 
However, these algorithms usually require large-scale high-quality annotated data, which is challenging to obtain.
Instead, the more practical semi-supervised setting has gained popularity~\cite{kim2016scalable, beitzel2005improving}, where unlabeled queries are used to boost the classification performance. When executed effectively, a well-performing hierarchical query classifier enhances the user experience, fostering a smoother and more productive interaction between users and the service.

However, building an accurate hierarchical query classification framework in real-world is non-trivial due to \underline{two challenges}: 
(1) \textit{severe class imbalance}. 
Take Amazon as an example, sensitive queries are infrequent, accounting for less than 0.05\%\textasciitilde0.15\% of all queries. Even worse, when training the classification model, only a small initial set of sensitive queries is accessible. This greatly hinders the development of a high-quality classifier.
(2) \textit{typically short and ambiguous query text in search queries}. The average search query is about three words~\cite{webpage2023}, leading to a weaker semantic understanding of queries for correct classification. 

To overcome these two challenges, we propose a semi-supervised machine learning framework utilizing the instance hierarchy and label hierarchy to enhance query representation learning and classification performance. Particularly, 

\underline{(1) for the class imbalance challenge}, we use contrastive learning to learn representations that attend to the minor classes through instance hierarchy. Intuitively, even if the number of queries under a child category can be small, the number of queries under the corresponding parent category is large and these queries are close to each other. To leverage them, we adopt contrastive learning where we create positive pairs from the queries under the same child category while negative pairs from cross-child categories. We formulate it as an intra-class hierarchy in instance hierarchy, and further extend it to the inter-class hierarchy, where we consider positives as queries across child categories and negatives as queries across parent categories, as shown in Figure~\ref{fig:proposed_pipeline}. This instance hierarchy helps capture fine-grained information for model training.

\underline{(2) for the ambiguous and short text challenge}, we utilize the information from three parts to enhance the understanding of a query — i.e., 
(i) the neighboring queries that are under the same child category. Motivated by the fact that not all queries are short and ambiguous, especially, some neighboring queries are clear and distinguishable, we can leverage this similarity between neighboring queries by the aforementioned intra-class hierarchical contrastive learning; 
(ii) the neighboring child categories that share the same parent category. The intuition is that when implicitly aligning the query to different child categories under the same parent category, the model learns the semantics of the query from other queries. This is achieved by adding a hierarchy-aware loss in our classification task; 
(iii) the text information in the label usually contains useful signals like the semantic meaning that contributes to the downstream classification. Based on this, we first adopt BERT~\cite{devlin2018bert} to encode the label text into an embedding vector to capture the contextualized semantic embedding. We concurrently create a ``label'' graph to take hierarchies/relationships between labels into account and employ the previously generated embedding vectors as node features for downstream graph representation learning. We finally combine the representation vector with the query embedding vector to form the finalized feature vector of a query for the final classification task. Since we use hierarchical label information in this step, we define this process as label hierarchy. Altogether, the designed instance hierarchy and label hierarchy components aim to address the aforementioned challenges for better classification.

In the real-world e-commerce search, we have \underline{one observation} that there exist many unlabeled queries that share the typographical similarity to the annotated queries of the same category due to potential typos. 
These queries, when used as training examples, potentially assist the classifier by improving the robustness against mis-typed queries, as they are typographically close to the queries to augment the dataset. 
Based on this finding, we deploy the self-training learning stage in our pipeline, i.e., we use the pseudo labels of classified queries to retrain the model. When selecting queries for self-training, inspired by the aforementioned observation, we develop neighborhood-aware sampling to effectively identify high-quality similar queries. 
We argue that the observation of topographical similarity can be extended to generic semantic similarity. The proposed self-training pipeline can be adapted as well. Besides, we can interpret this observation from the adversarial learning perspective where we adversarially generate similar queries or identify similar queries from our unlabeled queries for model training to improve the robustness of the classifier. Overall, we deploy the self-training in our framework to utilize the crucial unannotated queries for classification performance gain.

To evaluate the proposed method, we examine it on proprietary Amazon data and public Web of Science and RCV1-V2 datasets using Micro-F1 and Macro-F1 scores. Our proposed method achieves the best performance in most cases across all compared methods and datasets, except for Micro-F1 on Web of Science and RCV1-V2 dataset. 
However, Micro-F1 is less critical than Macro-F1 in real-world applications since we have an imbalanced class distribution and need to focus on minority classes, which is attended to by Macro-F1.
Our result demonstrates the efficacy of our proposed method, especially on the Amazon dataset. 
Our method is generalizable to solve hierarchical query classification tasks in all domains and paves the path toward the next generation of hierarchy-aware query classification. 
The main contributions of our work are: 
\begin{itemize}
    \item We propose a new algorithm that utilizes the instance and label hierarchy through contrastive learning-enhanced representation learning, which allows us to leverage hierarchical information in a fine-grained manner to improve classification performance.
    \item We propose a neighborhood-aware sampling technique to selectively choose high-quality unlabeled data points for self-training boost.
    \item Extensive experimental results on both proprietary and public datasets demonstrate the superiority of our proposed method in most cases.
\end{itemize}

\section{Related Works}
In this section, we briefly introduce relevant research areas.

\subsection{Hierarchical Query Classification}
Hierarchical query classification aims at classifying queries into a category within a given taxonomy to understand user intent and facilitate downstream recommendation tasks. It can be formulated as a text classification problem where the input text is a combination of short keywords~\cite{zhou2017survey}. Existing conventional methods employ either a single flattened multi-class classifier or multiple binary classifiers. Based on extracted query features, these works can be categorized into two groups: 
(1) N-gram-based features~\cite{cavnar1994n}: Since the query keywords are indicative of the category it belongs to, the count of keywords can serve as features; 
(2) Embedding-based features~\cite{devlin2018bert, pennington2014glove}: Due to the advances in deep learning and natural language processing, some researchers use word embeddings (e.g., Glove) and contextualized embeddings (e.g., BERT) to represent the query for classification. Later, researchers designed advanced classification models and learning diagrams utilizing additional information from queries to enhance classifiers. 
For instance, ~\citet{liu2019system} proposed a mixture of conventional neural network and Naive Bayes as a classifier~\cite{liu2019system} while ~\citet{wang2022incorporating} incorporated the hierarchy of label information by a graph encoder into the text encoder~\cite{wang2022incorporating}. 
Besides, the context-aware session information~\cite{cao2009context} and searcher engagement data~\cite{he2023hiercat} are explored as well. 
Different from these existing efforts relying on extra information or overlooking abundant unlabeled data, we aim to boost performance using only easily accessible query and label data combined with unlabeled queries.

\subsection{Imbalanced Learning}
Class imbalance is a common issue in text classification, especially prominent in the hierarchical setting~\cite{chawla2002smote}. 
To address it, one commonly-used solution is re-sampling, which involves oversampling the minority class, under-sampling the majority class, or combining both to achieve a balanced class distribution~\cite{longadge2013class}. Another strategy is cost-sensitive learning~\cite{thai2010cost}, where higher costs are assigned to the misclassification of minority classes during model training, eventually making the model more sensitive to the minority class. 
The Synthetic Minority Over-sampling Technique is another notable approach that generates synthetic instances of the minority class to balance the class distribution~\cite{chawla2002smote}. 
For instance, ~\citet{pereira2021toward} utilized the path and depth information to oversample and undersample data points to improve the classifier.
Different from the previous works, we utilize the unlabeled queries that are predicted as minority classes to augment datasets.

\subsection{Contrastive Learning}
Contrastive learning has emerged as a powerful paradigm in unsupervised and self-supervised learning techniques~\cite{jaiswal2020survey} by significantly reducing the performance gap between supervised and unsupervised learning. At its core, contrastive learning aims to learn similar representations for semantically similar instances and dissimilar representations for distinct ones. It accomplishes this by 
distinguishing the ``positive'' pairs (two similar data points) from the ``negative'' pairs (two dissimilar data points). 
Especially by leveraging large amounts of unlabeled data, it opens up new avenues for model training in scenarios where labeled data is scarce or expensive to obtain.
The effectiveness of this approach has been showcased in numerous applications, such as image, speech recognition, and natural language processing~\cite{le2020contrastive}.

\subsection{Semi-supervised Text Classification}
Semi-supervised learning is a promising research direction since it utilizes both the labeled and unlabeled data points in machine learning~\cite{van2020survey}, which alleviates the high cost of data annotation.
Among the semi-supervised learning techniques, self-training is widely used where a model is initially trained on a limited set of labeled data points, and then iteratively expands the training datasets by using classified unlabeled data points~\cite{ye2020zero}. 
Different selection methods are developed to choose which data points for augmentation, including probability-based and uncertainty-based solutions~\cite{amini2022self, tanha2017semi, amini2022self}

\section{Problem Definition}

In this section, we provide the mathematical definition of the hierarchical query classification problem. Note that, for the sake of simplicity in presentation, we assume the problem space is a two-level category hierarchy, but, the proposed method is extensible to accommodate a multi-level category hierarchy.

We have a set of queries $Q = \{ q_1, q_2, ..., q_N\}$, where $q_i$ is the $i$-th query. A query is a sequence of words, represented as 
$q_i = x_1, x_2, ..., x_i, ..., $ where $x_i$ is the $i$-th word in a query. 
We can divide the query set $Q$ into two groups: 
unlabeled queries $Q_U = \{ q_1^U, q_2^U, q_3^U, ... \}$ and labeled queries $Q_L = \{ q_1^L, q_2^L, q_3^L... \}$.
For one labeled query, we have its child category $c_k$ and parent category $p_j$ where $p_j$ denotes the parent category from a set of parent categories $P = \{ p_1, p_2, p_3, ... \}$ and each parent category $p_j$ consists of a set of child categories $p_j = \{ c_1, c_2, c_3, ... \}$. In this case, we have $c_k \in p_j$ 
The goal is to leverage the information in both labeled and unlabeled queries to learn a function $\mathcal{F}(q_i) \to {c_k,p_j}$, where $ q_i \in c_k, q_i \in p_j $, and $c_k \in p_j$.

\begin{figure*}[!t]
\centering
\includegraphics[width=0.85\linewidth]{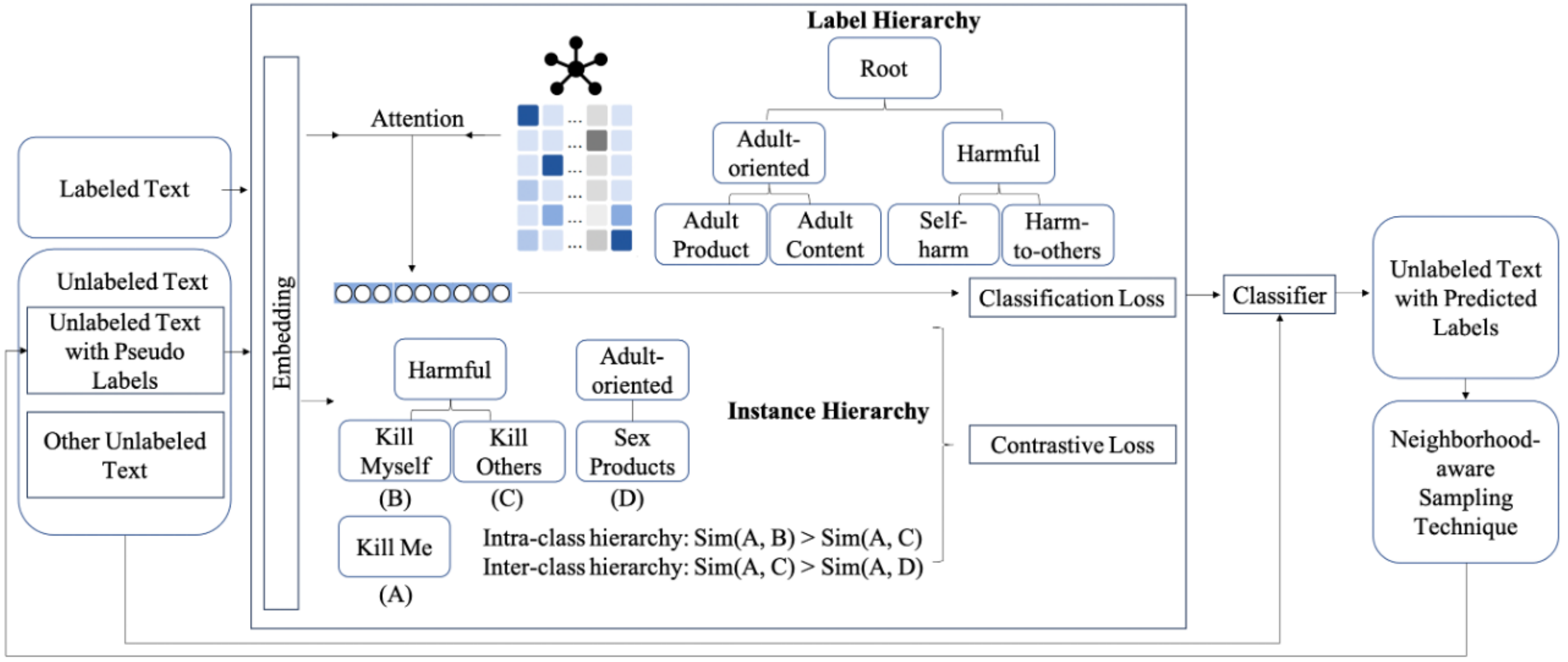} 
\caption{The overview of the proposed framework.} 
\label{fig:proposed_pipeline}
\end{figure*}

\section{Proposed Method}

In this section, we provide the details of the proposed semi-supervised hierarchical query classification framework to accurately classify a query for a given taxonomy. 
Specifically, the framework has three major components:
i) It utilizes the hierarchical label information to enhance initial query embeddings. 
ii) It attends to fine-grained instance hierarchy by modeling intra-class and inter-class relationships. The resultant contrastive loss boosts the query embedding learning, which is finally combined with a classification loss to train the classifier. 
iii) Through the proposed neighborhood-aware sampling technique, it selectively chooses high-quality unlabeled data points with pseudo labels to augment existing labeled data for model re-training. 
An overview of our proposed method is presented in Figure~\ref{fig:proposed_pipeline}.

\subsection{Label Hierarchy}

Given a query $q_i$, we pass it to BERT to get the textual embedding $emb_{q_i}$, following existing works to generate feature vectors~\cite{devlin2018bert, ma2023characterizing}. 
To attend to the hierarchy, we first create a label graph \( G=(V, E) \) representing the taxonomic hierarchy, where \( V \) is the vertex set of labels and \( E \) is the edge set of connections between parent and child labels. 
Because the label text (e.g., ``self-harm'', ``adult products'') contains useful semantic information for the downstream classification, we follow similar approaches to transfer text into embedding vectors~\cite{micallef2020role, he2021racism}, passing the label text to BERT to get the textual embedding as the node feature vector. 
For the root node, we use the average of all label embedding vectors. 
Due to the advances in graph neural networks for graph representation~\cite{zhou2020graph}, in practice, we pass the graph to a conventional two-layer graph convolutional network~\cite{kipf2016semi} to get the embedding as: 
\[
{emb}_G = \text{GCN}(G) 
\] 
Note that, not all label embeddings are equally relevant for a query. Motivated by the attention mechanism~\cite{vaswani2017attention}, we compute the attention score between a query and each label embedding. Formally, we have an attention matrix: 
\[ A_{q_i} = \text{Softmax}(({emb}_{q_i} \cdot W) \cdot {emb}_G^T) \] 
where $W$ is a matrix for feature dimension alignment between $emb_{q_i}$ and $emb_G$ during matrix multiplication. 
Then, we compute the attention-weighted label features, denoted as \[ emb_{q_i}^l = A_{q_i} \times {emb}_G .\] 
We finally concatenate it with query textual embedding \( {emb}_{q_i} \) derived by BERT and form the final representation \( {emb}_{q_i}^f = [{emb}_{q_i}, {emb}_{q_i}^l] \) for the downstream classification task. 
Particularly, for query \( q_i \), we predict its label as: 
\[\hat{y}_i = f({emb}_{q_i}^f). \] 

In addition, to leverage the hierarchical label information, we use the neighboring child label information to assist the classification by aligning the predicted child category to the neighboring child category. 
Such signal is incorporated into the model by a loss derived from between the current child category and the neighboring child category as: 
\[
\sum_{j \in p_i} \log(\hat{y}_j)
\] 
where \( p_i \) denotes the parent category of \( c_i \). 
By combing the two loss together, we have the final classification loss as:

\[
\mathcal{L}_{\text{Classification}} = \sum_{i}[y_i \cdot \log(\hat{y}_i) + (1-y_i) \cdot \log(1-\hat{y}_i) + \lambda \cdot \sum_{j \in p_i} \log(\hat{y}_j)]
\]
where $y_i$ is the label of the query and \( \lambda \) adjusts the importance of between the two losses. 
In implementation, we empirically set $\lambda = 1$. 

\subsection{Instance Hierarchy}
Given query \( q_i \), to learn its comprehensive representation in the context of its hierarchical structure, we use contrastive learning at two levels: (1) the intra-class hierarchy; and (2) the inter-class hierarchy. 
Particularly, for the intra-class level, we randomly sample two queries within the same child category as one positive pair \( (q_i, q_j), q_i \in c_i, q_j \in c_i \). For the query pairs that are in the same parent category but from different child categories, we treat them as negative pairs \( (q_i, q_k), q_i \in c_i, q_k \notin c_i, q_i \in p_k, q_j \in p_k \). The contrastive objective is defined as: 
\[ \mathcal{L}_{\text{intra-class}} = -\log \left( \frac{\exp(\text{sim}(emb_{q_i}, emb_{q_k}))}{\sum \exp(\text{sim}(emb_{q_i}, emb_{q_j}))} \right), \]
where \( \text{sim}(u, v) = \frac{u^Tv}{||u||_2 ||v||_2} \) denotes the cosine similarity between two vectors. Similarly, for the inter-class level, we have the predefined \( (q_i, q_k) \) query pair as the positive and the query pairs that are in different parent categories as negatives \( (q_i, q_l), q_i \in p_k, q_l \notin p_k \). The corresponding contrastive loss is: 
\[ \mathcal{L}_{\text{inter-class}} = -\log \left( \frac{\exp(\text{sim}(emb_{q_i}, emb_{q_l}))}{\sum \exp(\text{sim}(emb_{q_i}, emb_{q_k}))} \right). \]
Combining them together, the contrastive loss is given as: 
\[ \mathcal{L}_{\text{contrastive}} = w_{\text{intra-class}} \cdot \mathcal{L}_{\text{intra-class}} + (1 - w_{\text{intra-class}}) \cdot \mathcal{L}_{\text{inter-class}} \]
where \( w_{\text{intra-class}} \) denotes the weight of the intra-class hierarchy in the contrastive loss.

\subsection{Objective Function and Model Training}

Finally, we combine the classification and contrastive loss together, which arrives at:
\[
\mathcal{L} = w_{\text{contrastive}} \cdot \mathcal{L}_{\text{Contrastive}} + (1 - w_{\text{contrastive}}) \cdot \mathcal{L}_{\text{Classification}}
\]
where \( w_{\text{contrastive}} \) indicates the weight of contrastive loss in the final loss computation. 

When training the model, we minimize the loss for optimization through back-propagation using the Adam optimizer~\cite{kingma2014adam}.

\subsection{Neighborhood-aware Sampling}

After training, we apply the classifier \( \mathcal{F} \) to predict labels of the unlabeled data \( Q_U \) and then use the classified high-confidence data points to retrain the classifier. Motivated by our aforementioned observation that queries with similar labels tend to share similar typographical representations, we develop a neighborhood-aware sampling algorithm containing the following steps:

Given an unlabeled query \( q_{i}^U \in Q_U \), after inference by \( \mathcal{F}(q_{i}^U) \), we have the predicted child category and parent category: 
\[ \hat{c}_{q_{i}^U}, \hat{p}_{q_{i}^U} = \mathcal{F}(q_{i}^U) \].

\textbf{Step I:} We use the K-Nearest Neighbors (KNN) to find the labeled queries similar to the unlabeled queries in the feature space. Here, motivated by the aforementioned observation, the feature space can be the simple \( N \)-length string space and we compute Levenshtein Distance (Edit Distance) neighborhood search~\cite{berger2020levenshtein}. More generally, we use the previously generated BERT embedding to represent the feature space due to its powerful semantic representation in a broader case. Formally, we have:
\[\text{Neighbor}_{q_{i}^U} = \{ q_j | q_j \in \text{KNN}(q_{i}^U), q_j \in Q_L \} \] where \( q_j \) is from the labeled query set \( Q_L \) and its child category is \( c_j \) and parent category is \( p_j \). 
In practice, we utilize the hierarchical navigable small world method for the indexing and search process because of its efficiency in high-dimensional data spaces, making it a suitable choice for large-scale and high-dimensional datasets~\cite{malkov2018efficient}. 
To measure the similarity between queries, we adopt the cosine similarity metric. By using these two off-the-shelf solutions, we aim to achieve an efficient and accurate KNN search process.

\textbf{Step II:} After getting the neighboring labeled queries, we need to compute their distribution for the downstream sampling. To achieve this goal, we leverage the child category information between labeled and unlabeled queries.
The intuition is that if the unlabeled query shares the same child category of the labeled query, chances are high that this predicted child category is correct. 
Specifically, we compute KL divergence scores between the child category of the labeled query \( q_{j}^L \), denoted as \( c_{q_j^L} \), and the predicted child category of unlabeled query \( q_{i}^U \), denoted as $\hat{c}_{q_{i}^U}$. This score is denoted as KL distance:
\[ KL_{L-U_{\text{child}}} = \text{KL}(c_{q_j^L}, \hat{c}_{q_{i}^U}) \]
Since there are multiple labels in the neighborhood of unlabeled query $q_i^U$, it is necessary to take the quality of these queries into consideration during the sampling. 
Formally, for \( q_{j}^L \in \text{Neighbor}_{q_{i}^U} \), the average child category label is given as:
\[\bar{c}_{\text{Neighbor}_{q_{i}^U}} = \frac{1}{|\text{Neighbor}_{q_{i}^U}|} \sum c_{q_j}^L \]
We then compute the divergence score as:  
\[ KL_{L-L_{\text{child}}} = \text{KL}(c_{q_j^L}, \bar{c}_{\text{Neighbor}_{q_{i}^U}}) . \]
Finally, we add the scores from the child category information together as the final distribution:
\[
\text{Dist}(\text{Neighbor}_{q_{i}^U})_{\text{child}} = KL_{L-U_{\text{child}}} + KL_{L-L_{\text{child}}}
\]

\textbf{Step III:} Similarly, we compute the distribution based on the parent category information and get \\
\( KL_{L-U_{\text{parent}}} = \text{KL}(p_{q_j^L}, \hat{p}_{q_{i}^U}) \) and \( KL_{L-L_{\text{parent}}} = \text{KL}(p_{q_j^L}, \bar{p}_{\text{Neighbor}_{q_{i}^U}}) \), where \( \bar{p}_{\text{Neighbor}_{q_{i}^U}} = \frac{1}{|\text{Neighbor}_{q_{i}^U}|} \sum p_{q_j^L} \). After addition, we have 
\[
\text{Dist}(\text{Neighbor}_{q_{i}^U})_{\text{parent}} = KL_{L-U_{\text{parent}}} + KL_{L-L_{\text{parent}}} 
\]
Finally, we have the sampling distribution as: 
\begin{align*}
\text{Dist}(\text{Neighbor}_{q_{i}^U}) = & \, w_{\text{child}} \cdot \text{Dist}(\text{Neighbor}_{q_{i}^U})_{\text{child}} \\
& + (1 - w_{\text{child}}) \cdot \text{Dist}(\text{Neighbor}_{q_{i}^U})_{\text{parent}}
\end{align*}

\textbf{Step IV:} We sample the unlabeled queries following:  
\[
\text{Prob} \propto \text{Dist}(\text{Neighbor}_{q_{i}^U}) 
\]
We then add the sampled data points: 
\( 
\{(q_{i}^U, \hat{c}_{q_{i}^U}, \hat{p}_{q_{i}^U})\} \) to our existing labeled queries: 
\[ 
Q_L = Q_L + \{(q_{i}^U, \hat{c}_{q_{i}^U}, \hat{p}_{q_{i}^U} ) \} 
\] 
to retrain the classifier. 
After re-training, we run the neighborhood-aware sampling again to select new queries to augment the existing labeled datasets.
We repeat these steps until the model converges.

\section{Experimental Evaluation}

In this section, we examine the performance of the proposed framework by conducting extensive experiments.
Specifically, we aim to answer the following research questions:
\begin{itemize}[leftmargin=*,noitemsep]
    \item RQ1: How effective is our proposed method when compared to other methods?
    \item RQ2: What is the contribution of each component in the proposed framework?
    \item RQ3: How sensitive is the model performance when we change the parameters?
\end{itemize}

\subsection{Datasets}
We evaluate the proposed framework on both public and proprietary datasets.

\subsubsection{Public Datasets} We adopt two benchmark datasets widely used for hierarchical text classification, including Web-of-Science and ECV1-V2.  The data statistics is shown in Table~\ref{tab:pubic_data}.
\begin{itemize}[leftmargin=*,noitemsep]
    \item Web-of-Science (WoS)~\cite{kowsari2017hdltex}: 
    WOS dataset contains keywords and abstracts of academic papers across several disciplines, e.g., economy and science.
    For one paper, it also has a hierarchical domain-area label, representing the hierarchical nature of the discipline (e.g., Computer Vision domain under Computer Science category). This makes WOS suitable for the hierarchical query classification task where we treat keywords, area, and domain as query, child category, and parent category, respectively.

    \item RCV1-V2~\cite{lewis2004rcv1}: 
    The RCV1-V2 dataset is a benchmark corpus for text categorization research where each document has metadata such as date and title, in addition to the content of the news story. It comprises an archive of over 800,000 manually categorized newswire stories from Reuters Ltd. Its hierarchical categorization scheme includes four main topics (Corporate/Industrial, Economics, Government/Social, and Markets), which are further divided into subtopics, leading to over 100 leaf-level categories. Thus, it is an excellent dataset for hierarchical classification tasks. We use extracted nouns from titles, subtopic, and main topic as query, child category, and parent category, respectively.
\end{itemize}

\begin{table}[!t]
    \centering
    \caption{Data statistics of public datasets.}
    \label{tab:pubic_data}
    \begin{tabular}{|c|c|c|}
     \hline
         Public Dataset Name&  \# of Classese& \# of Hierarchy\\  \hline
         Web of Science (WoS) &  141& 2\\  \hline
         RCV1-V2&  103& 4\\  \hline
    \end{tabular}
\end{table}

\subsubsection{Proprietary Dataset}

For proprietary dataset, we use Amazon search queries as our testbed for examination on real-world application settings. We sampled 9\textasciitilde10 million user queries to create the dataset. 
In the Amazon dataset, a substantial portion consists of unlabeled queries, making up 40\% to 50\% of the total. Labeled non-sensitive queries also represent a significant segment, comprising 45\% to 58\% of the data. Within the labeled sensitive categories, queries related to adult-oriented products form 3\% to 6\%, while adult content is less common, constituting only 0.3\% to 0.5\%. The dataset also includes a small fraction of queries that are potentially harmful, with those related to self-harm and harm to others present in 0.003\% to 0.005\% and 0.01\% to 0.03\%, respectively.
The remaining sensitive queries count for 0.04\% to 0.07\%.

\subsection{Evaluation Setup}
\subsubsection{Evaluation Metrics} Since it is a standard imbalanced data classification task, we follow the existing related works and measurement: the Micro and Macro F1 score~\cite{wang2022incorporating, wang2022hpt}.
\subsubsection{Compared Methods} We compare with the standard multi-class text classifiers using fine-tuned BERT. Besides, several state-of-the-art (SOTA) hierarchical text classifiers using transfer learning and prompt learning are examined~\cite{wang2022incorporating, zhou2020hierarchy, banerjee2019hierarchical, khudabukhsh2015building, ma2021label}. 
Namely, we compare with 
(1) HPT~\cite{wang2022hpt}, where prompt tuning on pre-trained language model is utilized to handle hierarchical classification from a multi-label masked language model perspective; 
(2) HGCLR~\cite{wang2022incorporating}, where new queries will be generated by the label hierarchy to enhance the query representation learning using the contrastive loss; 
and (3) HiTIN~\cite{zhou2020hierarchy}, where the label hierarchy is converted into an unweighted tree structure to enhance the query representation. 

In implementation, to utilize the unlabeled queries for a fair comparison, we add the widely-used confidence-based sampling strategy in the self-training stage for each compared method. This ensures that all methods are compared in the same semi-supervised setting. 
For the train/val/testing split, we follow the existing split for the public Web of Science and RCV1-V2 datasets, and an 80-10-10 split for the Amazon dataset. 
For unlabeled data points, we use the existing unlabeled data points in the Amazon dataset and 10\% of the whole public Web of Science and RCV1-V2 datasets.

\subsection{Evaluation Results}

\subsubsection{RQ1: Effectiveness of the proposed framework}

As we see the comparison results in Table~\ref{tab:merged_res_results}, our proposed method is the best in most cases except Micro-F1 on Web of Science and RCV1-V2 dataset.

\begin{table}[!t]
    \centering
    \caption{Comparison of hierarchical query classification performance. The best algorithm in each row is colored in dark
{blue} and the second best is light blue. 
Note that we present the baseline result on Amazon as ``0'' for relative comparison (Here, the baseline is BERT), and $\pm$ indicates that the corresponding method is above or below the baseline.}
    \label{tab:merged_res_results}
    \begin{tabular}{|c|c|c|c|c|c|c|}
    \hline

Dataset & Metric & BERT & HPT & HGCLR & HiTIN & Ours \\ \hline
\multirow{2}{*}{Amazon} & Micro-F1 & 0 & -0.31 & +2.90 & \cellcolor{blue!10}+2.91 & \cellcolor{blue!25}+3.26 \\ 
& Macro-F1 & 0 & +0.67 & +2.56 & \cellcolor{blue!10}+3.35 & \cellcolor{blue!25}+4.10 \\ \hline

        \multirow{2}{*}{WoS} & Micro-F1 & 47.98 & 44.27 & 51.73 & \cellcolor{blue!25}54.21 & \cellcolor{blue!10}53.19 \\ 
        & Macro-F1 & 42.93 & 43.71 & \cellcolor{blue!10}48.92 & 47.93 & \cellcolor{blue!25}50.54 \\ \hline
        \multirow{2}{*}{RCV1-V2} & Micro-F1 & 74.96 & 73.07 & 76.57 & \cellcolor{blue!25}76.95 & \cellcolor{blue!10}76.91 \\ 
        & Macro-F1 & 58.49 & 58.78 & 59.25 & \cellcolor{blue!10}60.84 & \cellcolor{blue!25}61.48 \\ \hline
    \end{tabular}
\end{table}

This demonstrates the efficacy of our proposed method, especially on the Amazon dataset. 
In detail, we find that our proposed method beats the baseline fine-tuned BERT with the largest margin, indicating the necessity of designing sophisticated approaches for performance gain (i.e., instance hierarchy, label hierarchy, and neighborhood-aware sampling). 
We also beat the advanced HPT and HGCLR solutions. The reason may be that we explicitly consider the instance hierarchy to model the relationship between queries while HPT and HGCLR focus more on the hierarchical label structure. Our neighborhood-based sampling technique also contributes by selecting high-quality data points for self-training. 
Even if we are weaker than HiTIN regarding Micro-F1 on Web of Science and RCV1-V2, we are better in Macro-F1, which is more crucial.
This is because in the real-world application setting, like the sensitive query classification on Amazon, critical categories have fewer data points and we should treat each category equally rather than each data point equally during evaluation. This is achieved by the Macro-F1 score.

\subsubsection{RQ2: Ablation studies}

To examine the contribution of each component in the proposed framework (i.e., label hierarchy, instance hierarchy, and neighborhood-based sampling in the self-training stage), we first remove one component in the framework. Then, we retrain the model and measure the classification performance.

\begin{table}[!t]
    \centering
    \caption{Ablation studies of our proposed framework.
Note that we present our method on the Amazon dataset as ``0'' for relative comparison, and $\pm$ indicates that the corresponding ablated method is above or below our method.}
    \label{tab:merged_ablation}
    \begin{tabular}{|c|c|c|p{1.3cm}|p{1.3cm}|p{1.2cm}|} 
    \hline

Dataset & Metric & Ours & Ours w/o \newline label \newline hierarchy & Ours w/o \newline instance hierarchy & Ours w/o \newline self-training \\ \hline
\multirow{2}{*}{Amazon} & Micro-F1 & 0 & -4.92 & -1.56 & -1.06 \\ 
& Macro-F1 & 0 & -6.25 & -1.04 & -1.94 \\ \hline

         \multirow{2}{*}{WoS} & Micro-F1 & 53.19 & 48.17 & 51.41 & 52.27 \\ 
         & Macro-F1 & 50.54 & 47.95 & 48.02 & 49.16 \\ \hline
    \end{tabular}
\end{table}

As shown in 
Table~\ref{tab:merged_ablation}, our proposed method with all components is better than any revised method that is removing one component. This demonstrates the necessity of each component in the pipeline. 
Interestingly, we find removing the label hierarchy leads to the largest performance drop, possibly because the additional information from the label text shares certain similarities with the query text embedding, thereby contributing to the model training.
The instance hierarchy and self-training stage contribute to the performance to a similar degree. 

\subsubsection{RQ3: Hyperparameter tuning}

In this section, we examine the effect of hyperparameters on the model performance. Here, we focus on three major parameters $w_{\textrm{intra-class}}, w_{\textrm{contrastive}}$, and $w_{\textrm{child}}$.  The value ranges from 0.01 to 1 and we report the performance in 
Table~\ref{tab:merged_parameters_amazon}.

\begin{table}[!h]
    \centering
    \caption{Effects of varying weights on Amazon dataset.
Note that we present the result by our method (i.e., $w_{intra-class}=0.1$, $w_{contrastive}=0.01$, and $w_{child}=0.1$) on the Amazon dataset as the baseline, denoted as ``0'' for relative comparison, and $\pm$ indicates that the corresponding method configured with certain parameter values is above or below the compared method. $\Delta$ Micro-F1 means the difference in the Micro-F1 score and $\Delta$ Macro-F1 means the difference in the Micro-F1 score.
    }
    \label{tab:merged_parameters_amazon}
\begin{tabular}{|c|c|c|c|c|} 
\hline
Parameter & Value & $\Delta$ Micro-F1 & $\Delta$ Macro-F1 \\ \hline
\multirow{6}{*}{$w_{intra-class}$} & 0.1 & 0 & 0 \\ 
& 0.3 & -0.78 & -0.64 \\ 
& 0.5 & +0.10 & +1.67 \\ 
& 0.7 & +0.19 & +1.02 \\ 
& 0.9 & +0.62 & +2.66 \\ 
& 1 & +0.12 & +2.04 \\ \hline
\multirow{6}{*}{$w_{contrastive}$} & 0.01 & 0 & 0 \\ 
& 0.1 & +1.47 & +2.25 \\ 
& 0.3 & +0.95 & +0.98 \\ 
& 0.5 & -0.42 & -1.01 \\ 
& 0.7 & +0.28 & -2.64 \\ 
& 0.9 & -3.66 & -6.49 \\ \hline
\multirow{6}{*}{$w_{child}$} & 0.1 & 0 & 0 \\ 
& 0.3 & +0.69 & +0.25 \\ 
& 0.5 & +0.35 & -0.15 \\ 
& 0.7 & +0.24 & -0.19 \\ 
& 0.9 & +0.25 & -0.21 \\ 
& 1 & -0.14 & -0.34 \\ \hline
\end{tabular}
\end{table}

As we can see, 
(1) for $w_{\textrm{inter-class}}$, the best value is 0.9 and the higher value leads to better performance except 1, which indicates that the intra-class hierarchy contributes more than the inter-class hierarchy. The reason can be that intra-class helps the model learn the representation better for the downstream query classification; 
(2) for $w_{\textrm{contrastive}}$, we find when we add the contrastive loss to the classification loss, it helps improve the classification performance. But, the weight should not be large. 0.1 works best in the current setup.
(3) for $w_\textrm{{child}}$, we see that 0.3 is the best. It implies when utilizing both child and parent category information in the sampling stage, we should carefully choose and tune the weight. 
All these results indicate the contribution of each corresponding component. When using them together, we should take caution and exhaustively test different values to find the proper hyperparameter set for the specific application setting.

\section{Application in Practice}
We launched the proposed model in our sensitive query detection platform on Amazon, which is used in the search module. We compare the proposed method with our previous rule-based production model. We sample queries detected as positive by the model, and ask the human labeling team to measure precision before and after the launch. The results show that our method is better. 
\section{Conclusion and Limitation}

In this work, we propose a novel hierarchical query classification framework to effectively classify short queries into different groups. 
In essence, we first utilize label and instance hierarchy patterns to derive the classification and contrastive loss to train the model. 
We then design a neighborhood-aware sampling method to intelligently utilize unlabeled queries with pseudo labels to boost the model performance for self-training. Extensive results on both proprietary and public datasets demonstrate the effectiveness of our model. 

However, our work still suffers from a few limitations.
First, since our method is a multi-stage framework rather than an automatic end-to-end solution. Manual configuration and monitoring of the model training are needed, especially, during the self-training stage to determine the number of high-quality data points for the model retraining.
Second, in the use case of sensitive query classification, users can purposefully write queries to bypass or attack the classifier such that the classification performance drops~\cite{ebrahimi2017hotflip, he2021petgen}. We plan to explore this in future work.
Third, Large Language Models have gained popularity in recent years due to their promising results across multiple applications including text classification and text generation. Our method can beat ChatGPT during our preliminary examination of the Web of Science dataset. The potential reason is that our task is more challenging than the conventional text classification due to the complex label hierarchy structure. But, more efforts are required to accomplish a thorough comparison.
Fourth, our method still requires a large number of annotated data points. In the real-world application, especially the sensitive query classification on Amazon, there are only few annotated data points for certain categories. Few-shot learning can be a possible direction for future research.

\bibliographystyle{ACM-Reference-Format}
\balance
\bibliography{main.bib}

\appendix

\end{document}